# Improved prompting and process for writing user personas with LLMs, using qualitative interviews: Capturing behaviour and personality traits of users.

**Stefano De Paoli – s.depaoli@abertay.ac.uk**

**Abertay University**

## Abstract

This draft paper presents a workflow for creating User Personas with Large Language Models, using the results of a Thematic Analysis of qualitative interviews. The proposed workflow uses improved prompting and a larger pool of Themes, compared to previous work conducted by the author for the same task. This is possible due to the capabilities of a recently released LLM which allows the processing of 16 thousand tokens (GPT3.5-Turbo-16k) and also due to the possibility to offer a refined prompting for Personas' creation. The paper offers details of performing Phase 2 and 3 of Thematic Analysis, and then discusses the improved workflow for creating Personas. The paper also offers some reflections on the relationship between the proposed process and existing approaches to Personas such as the "data-driven" and "qualitative" Personas. Moreover, the paper offers reflections on the capacity of LLMs to capture user behaviours and personality traits, from the underlying dataset of qualitative interviews used for the analysis.

## Introduction

The goal of this draft paper is to build on work previously conducted by the author on performing a Thematic Analysis (TA) of qualitative data with a Large Language Model (LLM) (see De Paoli, 2023a and 2023b), following the 6 phases to TA proposed by Braun and Clarke (2006). In the previous work I showed it is possible to perform phases 2-5 of a TA (initial coding, generation of themes, themes validation and summarisation) using an LLM, with some degree of validity (De Paoli, 2023a). Moreover, I also showed that it is possible to perform phase 6 of a TA, which relates to the writing up of the results (De Paoli, 2023b). In relation to phase 6 specifically, in my previous work I proposed a workflow to write user personas with an LLM, based on the results of a TA of qualitative interviews also conducted with the support of the LLM. The previous work presented, however, a number of limitations, and it was in essence an initial attempt to perform phase 6 of a TA, whilst consolidating the phases 2-5. However, I believe there is scope to explore more the workflow for writing User Personas and attempt at producing far more articulated and composite Personas' narratives in phase 6. This is the goal of this manuscript.

The impetus for this work comes from recent developments in the areas of User Experience (UX) and LLMs. An online post which has captured the attention of the User Experience (UX) community, entitled "Can ChatGPT Replace UX Researchers? An Empirical Analysis of Comment Classifications", reported the results of performing a TA of a relatively small set of online user comments comparing the Themes produced by the LLM and by Human Analysts, showing overall a good agreement between the codes using the Cohen K measure (Schiavone et al., 2023). In part this confirms also my previous observations that, to an extent, it is possible to perform some form of TA with an LLM which show also some degree of inter-reliability between humans and LLMs. Most interestingly, the post was picked up by Jakob Nielsen, one of founding researchers in the field of usability, which commented that "this doesn't mean that ChatGPT can analyze user *behavior*, only that it is likely a time-saver in grouping non-behavioral questionnaire responses. (ChatGPT is known to be great at summarizing and classifying text without understanding what it means.)" [1] It is possible that Nielsen, when using the work 'behavior' was referring specifically to usability. Nonetheless it is an interesting problem to explore whether an LLM such as GPT3,5-Turbo could indeed capture some aspects of user's behaviours (or not) as part of the performing of a TA, and if we can reflect such behaviours in Personas, a key UX tool. It may actually be possible to test if a TA of user interviews done by an LLM can pick up aspects related with the user behaviour or the user personality traits, in order to build a user Persona. In my previous work, for building Personas I concentrated on needs and challenges only (i.e. what Nielsen probably call non-behavioural responses), largely for demonstrative purposes (i.e. demonstrating we could build some sort of textual output with LLMs, using the results of a data analysis). These two dimensions clearly did not encompass any specific behavioural trait or the personality aspects of a Persona.

The second factor giving impetus to this work is the recent release (July 2023) of a version of the LLM GPT3.5-turbo, which can accommodate a much large number of tokens (16 thousand) compared to the version I used in my previous work (which did accommodate 4097 tokens only). While the underlying logic of the larger LLM is exactly the same as the smaller one, the first then allows much larger prompts and responses, thus possibly allowing to work with more material as input and to produce much richer and extended textual narratives as output.

In the following pages I will first present some basic elements of the state-of-the-art on using LLMs for qualitative analysis or of coding of textual material, and of user personas and LLMs. I will then briefly present some key methodological aspects of this work, before turning to reflect on an improved process of building personas from a TA, done with an LLM. The data for this research is a set of user interviews with European farmers conducted by the H2020 project EUREKA (Vago and Spanoghe, 2023). The interview transcripts are available as open data from zenodo. The paper concludes with a discussion reflecting on the use of LLMs for building Personas and the relation to the broader field of UX research.

---

[1] https://www.linkedin.com/posts/jakobnielsenphd_can-chatgpt-replace-ux-researchers-an-empirical-activity-7067298800259760128-ZmsS/

# Literature review

In this section we will briefly discuss key aspects on the state of the art around doing qualitative analysis with the support of an LLM, and literature and other sources related to the use of LLMs in User Research and then focus on some key literature elements of User Personas.

To the author's knowledge there are 2 significant papers which have focused on using LLMs for qualitative analysis, at least at the level of initial coding and some other less developed or tangential contributions. In qualitative analysis coding normally comes early in the process (e.g. it is the second step of TA) and in essence it encompasses assigning a descriptive label to certain portion of data to capture specific aspects of the data, such as e.g. meaning, events. There are several strategies for coding and a very comprehensive overview is offered by Saldaña (2021). In relation to LLMs more specifically, Xhiao et al. (2023) and Gao et al. (2023) approached qualitative coding with the use of an LLM, showing good level of agreement between human codes and the LLMs, using Cohen's Kappa metric for inter-reliability. Gao et al. (2023) focused on collaboration in coding and proposed a tool called CollabCoder for conducting qualitative coding. Xhiao et al. (2023) offered interesting insights about the role of the prompt in the analysis, in relation to deductive coding with a pre-defined codebook. As described elsewhere my approach differs from theirs as I am working on having the LLM doing inductive TA, rather than working on deductive coding. Moreover, I sought to cover – at least for TA – the whole process of analysis, whereas Gao et al., 2023 and Xhiao et al., 2023 only focus on the coding phase of qualitative analysis. The inductive process of coding and the reuse of open data I am doing does not also immediately allow the use of inter-reliability metrics. Moreover, as I am using analysis performed by other scholars to validate my approach, it is not possible to guarantee that the coders (whether humans or LLMs), have seen the same portions of data in likely the same order. Therefore, in the first case (i.e. the works of Xhiao et al. and Gao et al.) the K metrics can be applied as the coding framework is decided in advance and there is a guarantee that the material is seen in the exact same form. In my case, no coding framework is decided in advance and as I work using chunks of data it is not possible to know if the LLM has seen the same data are the original coders, and the K metric cannot be used. However, I have shown with a comparative evaluation that an LLMs can generate most of the codes that human analysts produce, thus showing that an inductive TA done with an LLM, albeit still crude, has some degree of validity (De Paoli, 2023a). Moreover, I also conducted phase 6 of a TA, using the results of a TA to write intermediate research products like indeed user personas (De Paoli, 2023b).

Tangential work in the area of data analysis has seen experimentation with the use of LLMs for assisting with deductive content analysis (Chew et al., 2023), or for the annotation of data (Zhang et al. 2023), in particular audio-visual data. As anticipated, a recent online post by Schiavone et al. (2023) entitled "Can ChatGPT Replace UX Researchers? An Empirical Analysis of Comment Classification", has used chatGPT to conduct a TA for the classification of user comments, showing that human-to-human (k=0.7) and human-to-llm (k=0.63) inter-reliability of coding is very similar. Thus, further confirming previous observations that coding and even TA can be performed satisfactorily with LLMs, albeit in this case on a small sample of online comments.

Therefore, existing evidence preliminarily tells that performing a TA with LLMs is possible. This online post by Schiavone et al. (2023) is interesting also because it raises some observations about if and how LLMs will impact the UX research world. Some reflections on this were offered also in De Paoli (2023b), and it does seem that the conclusions are similar, namely that LLMs will not replace user researchers, but complement their work, and that UX researchers may need to acquire new skills to operate LLMs in their work.

This naturally leads, then, to consider the role of LLMs in the writing of user Personas as one of the possible tasks where UX researchers can rely on these new technologies. Personas are a key tool in UX research and are used as fictional but still realistic representations of users (Harley, 2015). They are fictional because they are narratives seeking to capture certain aspects of the target user groups. They are realistic because they are created from data collected from real users (e.g. see Cooper, 2007). Personas support designers in taking decisions, whilst avoiding just talking generically about "the user" and thus focusing on specific traits, behaviours, goals or frustrations of the real users. One of the benefits of using personas is that they can be a shared understanding tool for designers, developers, and stakeholders (Nielsen and Hansen, 2008). The most common type of personas used in UX work are built from qualitative data, most likely semi-structured interviews or ethnographic observations. The data analysis on e.g. interviews reveal patterns across the sampled users which can then be used as the building block of personas' narratives. TA is one of the approaches commonly used to analyse interviews for building personas (Rosala, 2019).

There also is an increasingly influential approach to building personas which is defined as "data-driven" (see e.g. Salminen et al., 2021, for a review; McGinn and Kotamraju 2008, for an early paper; Jensen at al., 2022 for a comprehensive overview; and Salminen et al. 2020 for reflections around ethics and bias in data driven personas), which relies on big data, analytics and algorithms. The name "data-driven" is to an extent misleading as also the personas built from rich qualitative data are data driven, but in essence "data driven personas" encapsulate the use of algorithms for analysing a much larger pool of often numerical data, such as previous surveys, demographics data, analytics, market research and so on. These personas can be written entirely by algorithms or with algorithms assisted by humans (Jansen et al., 2020). A specific subset of these has been called "automatic persona generation" which, beyond "data driven" personas, also see the personas narrative as automatically generated entirely by algorithms (Salaminen et al., 2019).

To an extent "data-driven" personas, could be seen has having some similarities with the approach I proposed in De Paoli (2023b) as the personas are built with the support of the LLMs algorithms. However, the personas written with LLMs that I am proposing are based on qualitative data analysis, rather than on quantitative data and related analysis. Consequently, they are closer to the "traditional" types of personas built from rich qualitative material. Moreover, the personas produced with LLMs have to be seen as the product of a Human-AI collaboration, as they are not entirely written by algorithms. Perhaps, one could see then the proposition to build personas with an LLM as a middle ground approach in-between the entirely qualitative personas, hand crafted by researchers, and the "data-driven/automatic" personas, produced with substantial work delegated to algorithms and relying on big data.

Lastly, the recent hype around LLMs has also seen attempts to build Personas with model applications such as e.g. chatGPT, with no underlying empirical data, and effectively reusing the training material of LLMs. However, in essence, the existing attempts are using LLMs to build entirely fictional Personas, not based on any actual data. These attempts can be found in several online posts and have for example been dubbed as synthetic personas[2]. The issues with creating entirely fictional Personas rest with the fact that they are not indeed based on using actual user data to derive insights that can help with e.g. system design. In my previous work I adopted a definition from Harley (2015) which states that Personas are "fictional, yet realistic" user representations. This, I believe is an important metric for building Personas as it can be used as the testing element of LLMs generated Personas. Indeed, those personas created with LLMs, without any underlying user data are fictional, but not realistic as they are not based on actual user data. On the other hand, LLMs generated Personas which are based on existing data and an analysis do also have the property of being realistic. Table 1 provides a brief overview of the types of Personas discussed above, in relation to both the kind of data upon which they are built and in relation to who writes the narratives. The focus of my work is on LLMs qualitative personas.

| **Personas Type** | Qualitative Personas | Data Driven Personas | LLMs Qualitative Personas | LLMs Fictional Personas |
|---|---|---|---|---|
| **Type of Data** | Qualitative (e.g. interviews) | Largely quantitative (e.g. survey analytics) | Qualitative (e.g. interviews) | No empirical data |
| **Written by** | Human | Automation (algorithms), sometimes human assisted | Human-AI Collaboration | AI (I.e. LLMs) |

**Table 1** – Brief summary of some examples of personas based on type of data used and who writes them

## Material and methods

For the purposes of conducting the Thematic Analysis I used the LLM GPT3.5-Turbo-16k[3]. This is a model that allows processing 16 thousand tokens (which cover both the input prompt and the output response). The model used in the previous work on Personas (see De Paoli, 2023b) allowed to use 4097 tokens during processing. Therefore, the 16k model supports a much larger prompting/response and would support using more material from an analysis for building

---
[2] https://niloufars.substack.com/p/i-tried-out-syntheticusers-so-you
[3] See https://platform.openai.com/docs/models

Personas, whilst at the same time allowing to produce a much more composite narrative. The model was accessed through the OpenAI API using ad-hoc python scripts, where the prompt to the model is part of scripts which perform additional operations on the data such as formatting, storing, or slicing. More details on the use of the API and python for performing the analysis can be seen from the previous papers (De Paoli, 2023a and 2023b) and will not be repeated here. I will instead concentrate on discussing some aspects of the prompts used.

A prompt is in essence the set of instructions we request the LLM to perform, and the creation of a prompt capable of producing the expected results is often referred to as 'prompt engineering'. For an overview of the concept and a catalog of prompting strategies, please see White et al. (2023). The production and refining of prompts are fundamental for having LLMs perform their tasks. In the case of this work, the prompting is the outcome of several testing until the desired output is reached, that is testing with different instructions how the LLM performs the task, until a suitable prompt is identified. As a TA has already been performed, the prompts used in this paper have already been tested and proved to work. However, some variations were introduced, which will be addressed below. This paper will perform a further TA on the open dataset of farmers from the EUREKA project (Vago and Spanoghe, 2023), already used in the previous work on writing personas with LLMs (De Paoli. 2023b). The methodological approach here repeats roughly the same steps, however there are some differences which will also be discussed below.

First the approach I propose seeks to perform a TA with an LLM following the key phases to a TA proposed by Braun and Clarke (2006) in their seminal work. The idea is to perform each phase in turn to arrive at a consolidated set of themes. After this it will be possible to perform the writing-up phase of the personas, that corresponds to phase 6 of TA. In particular, phases 2-6 can be performed alongside the LLMs (from the initial coding, up to writing). Phase 1, the familiarisation with the data, is done manually by the human researcher, largely to clean the data for processing (for example by removing introductory sections of the interview transcripts where the interviewer explains to the participant what the goal of the project is etc.) and for preparing chunks for analysis (this work created 30 chunks from the 13 interview of the dataset, ranging from about 600 words to 1800 words). Even if the new model accommodates 16k tokens, it is important to divide each interview into smaller chunks, this is because it facilitates processing, and it also allows to reach some degree of granularity in the generation of codes.

For the purpose of this paper, sets of Themes have been generated (through performing phases 2-3 of a TA) in this research covering specifically the following sets: users' goals, frustrations, behaviours and personality traits. Key for the generation of these sets of Themes is however the code generation, covered by phase 2 of a TA. I generated four sets of codes using the process and prompt shown in Figure 1, as the basis to build the four sets.

Each chunk of interview then was processed with the prompt (inside a for loop), for the identification of codes (see Figure 1). This, following the process detailed in De Paoli (2023a and 2023b), amounts to the beginning of phase 2 of a TA. In the prompt I offered the model a simple, but clear definition of what was sought from the data. In Figure 1 for example, there is a definition of what are "user personality traits" and then the request in the prompt to identify "up to 4" user personality traits. Lastly a final component of the prompt is the request to format the response as a json file, with the content (i.e. codes and quotes) all contained in one main key (e.g. 'traits'), which will support the processing of the code in the subsequent phases of the work.

```
for i in range(l):

    text = df.loc[i]['Interview_chunk']

    prompt = f"""

    User personality traits refer to typical patterns of behavior, thoughts, and emotions of Users.

    Identify up to 4 relevant user behaviours of the interviewee in the text below, provide a name for each behaviour, a summary description of the behaviour and a quote from the respondent for each behaviour no longer than 4 lines

    Format the response as a json file keeping names, descriptions and quotes together in the json, and keep them together in 'traits'.

    ```{text}```
    """
```

**Figure 1** – Prompt used for the generation of initial codes

Table two shows the definition used in each prompting to identify the four set of initial codes. I selected and/or adapted definitions retrieved directly from ChatGPT[4] (e.g. I asked ChatGPT to give me a simple definition of user goals and then reused with minimal changes into the prompt). Other definitions could be used, including from academic or practitioners' work. However, I wanted to work with very simple and direct definitions, and definitions given by ChatGPT are sufficient for this.

| Themes | Definition used in initial coding prompt |
|---|---|
| User behaviour | User behavior are the actions, interactions, and communications made by individuals while engaging with digital or physical products, or services |
| Frustration | Frustrations are feelings of dissatisfaction, resulting from obstacles, unmet expectations, or barriers to achieving one's goals. |
| Goals | User goals are the specific objectives or intentions that individuals aim to accomplish when interacting with a product or service. |
| Personality traits | User personality traits refer to typical patterns of behavior, thoughts, and emotions of users. |

Table 2 – Definitions used in the prompts for initial coding

---

[4] https://chat.openai.com/

The use of the four definitions in the prompt (as per Figure 1), thus led to the generation of four sets of initial codes. I asked the model to generate up to 4 codes for each chunk. The number of codes to be generated is set by the researcher. This number should probably be a good balance between granularity and avoidance of reaching the maximum number of tokens error of the model. To note that in this instance the initial coding is more granular than the one proposed in previous work on the same dataset (De Paoli, 2023), since we are working with smaller chunks of the interviews (max 1800 tokens, compared to about 2500 in previous work), and are asking the model to infer 4 codes, rather than 3 from each chunk. This is possible because the model used here accommodates 16k tokens.

As the LLM sees chunks of interviews separately (with no memory of what it has done previously), we have already observed in previous work that codes can be repeated (see De Paoli, 2023a) across the chunks. Therefore, I also operated a codebook reduction to detect duplicates and merge them with very similar/identical codes, whilst keeping track of the relative interview and quotes from which the code was initially generated. Ultimately this led to the generation of four sets of unique initial codes as detailed in the table below.

| Set of codes | Number of codes (after reduction of codebook) |
|---|---|
| Behaviour | 113 |
| Frustrations | 104 |
| Goals | 79 |
| Personality Traits | 65 |

Table 3 – Unique codes for each set

Following the establishment of the initial codes, it is then possible to approach phase 3 of a TA, which requires sorting and organising codes into Themes. Themes are the main outcome of a Thematic Analysis, and they effectively constitute patterns across the data. Due to the demonstrative nature of this paper which concentrates on advanced prompting for building Personas, I did not perform phase 4 and 5 of TA, as in essence the results from phase 3 are enough material (Themes) for testing the creation of Personas. Anyway, a suggested methodological approach for performing these two phases (4/5) with LLMs is described in De Paoli (2023a and 2023b), with the second paper also using the same dataset used here. Phase 4 and 5 are essential for producing a high-quality TA.

For creating Themes from the initial codes, I used the same prompt proposed in previous work which is shown below (Figure 2). This is a tested prompt which produces Themes by grouping codes, without giving any indication as to what the focus of Themes should be, in a completely inductive fashion. I instructed the model to generate 11 Themes for Behaviour, Frustrations and Goals, and 8 for Personality Traits (due to the lower number of unique codes). As discussed in previous work, the choice of how many Themes to generate is somewhat arbitrary and depends on the choices of researchers. I decided on 11 (and 8) just to have sufficient variety (even though this may create overlaps), but it is certainly possible to create a smaller number of Themes. Themes obtained as response to the prompt from the model were also formatted as a json (using another prompt) for processing in the write-up phase.

> *prompt = f"""*
>
> *Determine how all the topics in the following list of topics can be grouped together, and topics can also be in more than one group. \\*
>
> *Group all the topics numbers only and provide a name and a description for each group\\*
>
> *Create 8 significant groups\\*
>
> *Display the full list*
>
> *List of topics: {", ".join(topic_list)}*
>
> *"""*

**Figure 2 –** Prompt used for grouping and sorting codes into Themes

Once the Themes were generated, I approached the phase of writing up the Personas. I devised a python script making a random selection of two Themes from each set (i.e. 4 tuples each with 2 elements), which were then consecutively passed into a prompt for the LLM to write up a Persona based on the analysis. To note that each Theme passed to the model comprises: the Theme name, its description, all the underlying codes (and their description) and all the related quotes. Moreover a few of the Themes resulted in being quite large and I further broke them down into sub-groups to facilitate the production of the prompt, whilst limiting therefore the risk of reaching the maximum number of tokens (even though some Themes combinations would still lead to the error). The workflow, with the random generation of tuples is described in Figure 4. The new prompt used for generating the Personas is presented in Figure 3 and requires some additional explanations. The 4 sets (i.e. lists of codes composing a Theme) of tuples are passed in the prompt, clearly instructing the LLM about the content of each list (e.g. Frustrations, Goals), which are at the bottom of the prompt itself. These four sets are the material from which the LLM will build the Personas.

The prompt instructs the LLM to tell what the Persona basic demographics are including the name, age and country. Following, the prompt instructs the LLM to identify what the Persona main goal (taken from the list) is, and what is the Persona main frustration. Next the prompt asks the model to include a quote illustrating the Persona goal (also taken from the data shown to the model) and a narrative background (for bout 250 words). Further, the prompt instructs the model to include in the narrative two additional sub-goals and sub-frustrations, as bullet points. Lastly the prompt instructs the model to provide as bullet points lists of 3 behaviours and personality traits.

```
prompt = f"""
Using the lists provided below, write a user persona
The persona should be structured as follows:

Name and surname (realistic also based on the country)
Country: name of the country of the Persona (which must be European)
Age: Based on the lists tell if the persona is young, middle or old age
"Goal & Frustration": tell what the persona main goal (max 1) is and what is the persona main frustration, include a quote taken from the quotes in the goal list representing the main goal
"Narrative": include also a narrative background of the persona (max 250 words)

"Additional goals": identify with bullet points 2 additional goals (max 20 words each) and two additional frustrations (max 20 words each) of the persona.

"Behaviour": identify the personas key behaviours (max 3, 20 words each),

"Personality": identify the main persona personality traits (max 3, 20 words each).

List of frustrations: {", ".join(frustrations_list)}
List of goals: {", ".join(goals_list)}
List of behaviours: {", ".join(behaviours_list)}
List of personality traits: {", ".join(traits_list)}
"""
```

**Figure 3** – Prompt used for the generation of Personas using the Themes

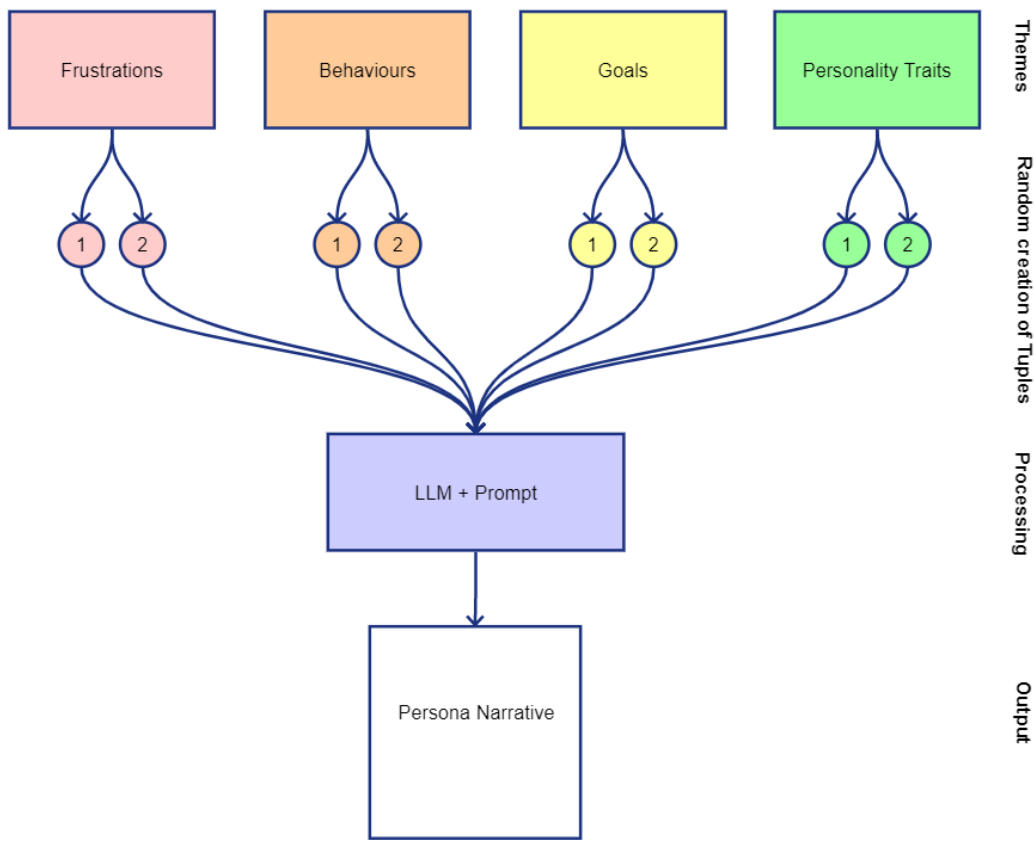

**Figure 4** – Workflow used for the generation of Personas, using the Themes

To note that occasionally the model would still produce a max tokens number error (i.e. the number of tokens processed, including prompt and response, was more than 16k), depending on the combination of themes used. I did not address this error in my scripts, despite taking precautions (like diving in parts the largest Themes from phase 3) as I am just trying to demonstrate an improved prompting and process for building personas. Nonetheless if the proposed process was to go into production it would be necessary to ensure that the length of the combination of Themes is not too big so as to produce the error, of which an example is presented in Figure 4.

```
InvalidRequestError: This model's maximum context length is 16385 tokens. However, your messages resulted in 19302 tokens. Please reduce the length of the messages.
```

**Figure 5** – Max Tokens error still produced occasionally, depending on combination of Themes

# Results of Thematic Analysis

This section presents the results of this work. We will start considering some elements of the Thematic Analysis and in particular examples of codes and the full list of Themes generated by the model, before looking at some examples of Personas generated by the LLM from the analysis.

## Codes

Phase 2 of TA relates to the initial generation of codes. In the approach discussed here as described earlier I asked the model to generate four sets of initial codes from the same dataset. Each set is directed by a simple definition of what the codes should be about. Some examples of the generated codes are shown below, for illustrative purposes and for providing an idea about the structure of the codes. We can see from these examples that the model does reasonably well in identifying the specific user aspects. Behaviours encompass for instance the need to validate in the field information found online, or the user/farmer seeking out others for collaboration. Personality traits encompass for instance persistence or the user being open-minded to innovation. Goals include for instance the need to obtain information online that can be trusted or information which clearly explains agriculture regulations. Lastly Frustrations encompass aspects such as finding unreliable information online, or dislike of websites with a lot of adverts.

| Set/Codes Example | Code Example | Description | Quote |
|---|---|---|---|
| **Behaviours** | Validation | The user uses digital tools to locate problems but validates them by going to the field to ensure that the problem is correctly identified. | But then I also go to the field to to validate what is a problem, because obviously digital toll can you tell you a lot but sometimes it's not enough. |
| | Problem-solving online | The behaviour of seeking out solutions to specific problems online, such as using a thermal camera to monitor cows during calving. | I looked online at what other options were available. A thermal camera, for example, was one of the options. A cow's temperature raises just before she calves. You can use this heat camera to measure the bodily temperature. |
| | Professional collaboration | The behaviour of users collaborating with advocacy and professional organizations through a given product | It is possible to connect advocacy and various professional organizations and everyone in the same way. There is no problem with that. |
| | Persistence | The ability to continue working towards a goal despite obstacles or challenges | The really small businesses with an unclear name. So I also worked a lot with raindance, that's a good |
| | | | name. That's something that you can not mistake. |
| **Personality traits** | Visionary | The interviewee is forward-thinking and believes that digital tools should be basic for everyone, and that future tools should integrate feedback from consumers and processors to provide guarantees about the way food is produced and processed. | I'm already beyond that. But this is kind of basic for everyone. So I I hope that all of those tools will be made more easy to accept by most farmers, by the 80 percent who could actually use them. And maybe today's only 20 percent who are already using them. So so there is still some some progress to be made there in the interface so that it's more user friendly...I believe tomorrow's tool will not only integrate on farm level but also the food chain and provides some guarantees somewhere to say yes, indeed this food was produced and processed by those companies and those farmers in such and such conditions. |
| | Open-mindedness | The willingness to consider new ideas and perspectives. | It is good to have the tool as complex as possible. Good experience with new tools. |
| **Goals** | Contacting businesses | The interviewee's goal is to contact businesses that produce sensors for agriculture through phone, email, and contact forms on their websites. | And then just to come to got to that contact forms, by phone, I prefer. But also with the e-mails and stuff like that. |
| | Trustworthy Information | The interviewee needs trustworthy information related to parcels, forest expansion, and manure. | We need trustable information, information on the parcels, to become forest or not, and manure (fertilizer, slurry, and solid manure). |
| | Clear Explanation of Regulations | The interviewee needs a clear explanation of regulations related to the expansion of forest plots, manure balance, and other farm-related issues. | Clear explanation (lay language) on everything what is needed for the farm and also next to the farm, especially about the expansion of forest plots and regulatory issues, manure balance. |
| | Dislike of search | The interviewee dislikes search engines with advertising, finding | The search engines that are sponsored by whatever. That's why I don't like the Mozillas and all of |
| **Frustrations** | engines with advertising | them less clear than Google. | those other stuff, because they are always, there is a lot of advertising on. |
| | Incomplete Tools | Frustration with digital tools that are incomplete and cannot fulfill all desired functions | I had some tools which were incomplete somehow, but that could not fulfill all of that I wanted from them. |
| | Unreliable Information | The interviewee expresses frustration with unreliable information. | Often information that is given cannot be trusted e.g. information given by 'Natuur en Bos' (Dpt of the Flemish Government), they give information on which birds you can shoot, who are damaging the cornfields (=fodder for the cattle), is not to be trusted. |

Table 4 - Example of initial codes generated by the model

**Themes**

The following tables present all the Themes generated by the model using the four set of codes, as input in the prompt presented in Figure 2. The Themes and the related description are presented in the same order as produced by the model and with the verbatim output (i.e. no change to the text was made by the author, for example the word group is used instead of Theme in the description as this was the instruction in the prompt which the model has reused).

| Behaviour | |
|---|---|
| **Theme** | **Description** |
| Online Information Access and Behavior | This group includes topics related to the behavior of accessing information online, such as searching for information, using digital tools, and filtering out misleading information. It also includes topics related to the preference for practical and illustrated content, as well as the importance of trustworthy and reliable online sources. |
| Offline Information Exchange and Collaboration | This group includes topics related to the behavior of exchanging information with colleagues offline, attending meetings and conferences, and collaborating with other farmers, researchers, and advisors. It also includes topics related to the preference for offline sources of information, such as journals and local meetings. |
| Language Preferences and Information Source | This group includes topics related to language preferences for accessing information online, as well as the impact of answer depth and information source on language choice. It also includes topics related to the use of different sources of information, such as scientific institutions and social media platforms. |

| Farm Management and Monitoring | This group includes topics related to the use of digital tools for farm management and monitoring, such as tracking animal fertility, controlling processes, and measuring aspects of work that cannot be measured with the naked eye. |
|---|---|
| Trust in Information Sources | This group includes topics related to the level of trust in different information sources, such as online sources, scientific institutions, and social media platforms. It also includes topics related to the preference for more reliable sources of information and the skepticism towards manufacturer claims. |
| Collaboration and Information Sharing | This group includes topics related to the behavior of collaborating with colleagues, exchanging experiences, and sharing information in forums and professional organizations. It also includes topics related to the willingness to share problems and collaborate with others in the same field. |
| Preferences for Visual Learning and Content | This group includes topics related to the preference for visual learning and content, such as videos and diagrams. It also includes topics related to the use of YouTube videos as a reliable source of information. |
| Preferences for Traditional Methods and Analog Communication | This group includes topics related to the preference for traditional methods of communication and problem-solving over digital ones. It also includes topics related to the value placed on practical experience, learning from older beekeepers, and the input of experts in the field. |
| Importance of Reliable and User-Friendly Digital Tools | This group includes topics related to the importance of reliable and user-friendly digital tools in agriculture. It also includes topics related to the desire for personalized and intuitive interfaces, as well as the emphasis on profitability and environmental impact. |
| Information Gathering and Problem-Solving | This group includes topics related to the process of gathering information and solving professional problems, both online and offline. It also includes topics related to the use of digital tools to solve professional problems and the reliance on practical experience and learning from others. |
| Use of Digital Tools for Specific Purposes | This group includes topics related to the use of digital tools for specific purposes in agriculture, such as beekeeping, farm management, and seeking information on agricultural machinery and equipment. |

Table 5 – Themes: Behaviour

| **Personality Traits** | |
|---|---|
| **Theme** | **Description** |
| Digital Savvy | This group includes topics related to the use of digital tools and technologies, including digital skills, using online resources, and seeking information through Google and other digital platforms. |
| Problem-solving Skills | This group includes topics related to the ability to analyze and evaluate information to make informed decisions and solve problems. |
| Attention to Detail | This group includes topics related to being meticulous and thorough in attending to details, paying close attention to small details, and ensuring accuracy. |
| Collaboration and Information Sharing | This group includes topics related to valuing working effectively with others towards a common goal and sharing knowledge and expertise. |
| Sustainability | This group includes topics related to concern for the environment and finding sustainable practices to reduce carbon emissions. |
| Willingness to Learn and Adapt | This group includes topics related to the desire to learn and explore new information and knowledge, being open to new ideas and perspectives, and adapting to new technologies and methods. |
| Traditional Values | This group includes topics related to valuing traditional methods and relationships in the agricultural world, social interaction, and relationships. |

Table 6 – Themes: Personality Traits

| Goals | |
|---|---|
| **Theme** | **Description** |
| Farming Information and Knowledge Access | This group focuses on topics related to accessing information and knowledge about farming practices, regulations, technologies, and innovations. It includes topics about staying informed, accessing trustworthy information, learning from industry professionals, and utilizing digital tools for farming operations. |
| Digital Tools for Farm Management | This group focuses on topics related to the use of digital tools for farm management, including data management, monitoring, problem-solving, decision-making, and communication. It includes topics about the development and usability of digital tools, as well as the benefits and challenges of using them. |
| Accessible and User-Friendly Digital Tools | This group focuses on topics related to the accessibility and user-friendliness of digital tools for farmers. It includes topics about smartphone compatibility, customization options, overview of important data, ease of use, and support systems for farmers using digital tools. |
| Agricultural Education and Training | This group focuses on topics related to incorporating digital tools into agricultural education and training. It includes topics about vocational training, online courses, and platforms for learning and obtaining licenses. |
| Precision Agriculture and Crop Management | This group focuses on topics related to precision agriculture and crop management using digital tools. It includes topics about real-time monitoring, early detection of crop issues, data validation, and reducing the use of chemicals. |
| Language and Localization | This group focuses on topics related to language and localization in digital tools for agriculture. It includes topics about providing content in local languages, avoiding the need for learning new languages, and focusing on agronomy as the core business. |

| Animal Breeding and Livestock Management | This group focuses on topics related to animal breeding and livestock management using digital tools. It includes topics about improving animal breeding processes, monitoring flock health, and utilizing online resources for livestock management. |
|---|---|
| Organic Farming and Sustainability | This group focuses on topics related to organic farming, sustainability, and conservation. It includes topics about staying informed about regulations and practices, implementing good agricultural practices, and exploring possibilities for compensation. |
| Agricultural Market and Competitiveness | This group focuses on topics related to the agricultural market and competitiveness. It includes topics about using digital tools for market feedback, testing new ideas and innovations, and staying ahead of the competition. |
| Information Verification and Reliability | This group focuses on topics related to verifying and ensuring the reliability of information in the agricultural field. It includes topics about filtering information, trusting digital knowledge from reliable sources, and verifying information found online. |
| Specific Farming Practices and Topics | This group focuses on topics related to specific farming practices and topics. It includes topics about machinery, crop issues, regulations, animal breeding, and managing fauna in the area. |

Table 7 – Themes: Goals

| **Frustrations** | |
|---|---|
| **Theme** | **Description** |
| Frustration with Digital Tools and Technology | This group includes topics that express frustration with difficulties in accessing information online, such as network availability issues, poorly maintained websites, language barriers, and the abundance of irrelevant or misleading information. It also includes frustrations with the lack of reliable sources, the need to verify information, and the difficulty in finding specific information or solutions to problems. |
| Frustration with Information and Knowledge | This group includes topics that express frustration with the lack of reliable and trustworthy information, difficulties in finding specific information, and the need for clear explanations and solutions. It also includes frustrations with the lack of collaboration, limited access to relevant magazines, and the lack of information on specific topics or practices. |
| Frustration with Agricultural Practices and Regulations | This group includes topics that express frustration with agricultural practices and regulations, such as difficulties in calculating feed rations, expanding forest plots, and filling production gaps. It also includes frustrations with the lack of clear information, politicized information, and the uncertainty of compensation for good practices. |
| Frustration with Beekeeping | This group includes topics that express frustration specifically related to beekeeping, such as difficulties in accessing education, relying on practical experience, lack of trust in digital sources, and the need for necessary digital skills. It also includes frustrations with irrelevant or inaccurate information and the lack of complete solutions from digital tools. |
| Frustration with Language Barriers | This group includes topics that express frustration with language barriers, such as difficulties in accessing or understanding information online due to language limitations. |
| Frustration with Lack of Visual Aids | This group includes topics that express frustration with the lack of visual aids, such as photos and videos, to aid in identifying diseases and other issues in farming. |
| Frustration with Limited Access to Information | This group includes topics that express frustration with limited access to information, such as offline sources being relied upon due to limited availability, lack of a central platform for agriculture information, and the small market and lack of big companies in certain regions. |
| Frustration with Lack of Training and Education | This group includes topics that express frustration with the lack of training and education, such as not having completed a particular training or education, difficulties in using digital tools due to lack of training, and the lack of clear vision for the role of digital tools in the future of agriculture. |
| Frustration with Information Overload | This group includes topics that express frustration with the abundance of information online, such as sifting through irrelevant or politicized information, the need to verify information, and feeling overwhelmed by the amount of information available. |
| Frustration with Communication and Collaboration | This group includes topics that express frustration with communication and collaboration, such as difficulties in personalizing digital tools, accessing important data, spreading information about events and programs, and the lack of user-generated content and exchange of information in the agricultural world. |

Table 8 – Themes: Frustrations

**Write-up: Personas**

In the following, four examples of Personas generated by the model with the prompt of Figure 3 and using the proposed workflow (Figure 4) and the Themes, are presented. This in effect amounts to Phase 6 of TA. The list of the underlying Themes to each Persona is also provided. Pictures of Personas were generated with Dall-e using part of the Personas narrative. Further examples of Personas are included in the appendix for reference.

| **Persona – Example 1** | |
|---|---|
| **Name and surname:** Sofia Müller<br>**Country:** Germany<br>**Age:** Young age | 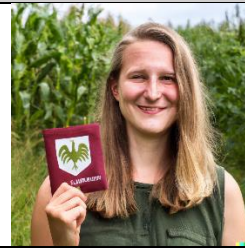 |

**Goal & Frustration:** Sofia's main goal is to access trustworthy information related to her farming activities. Her main frustration is the lack of reliable sources of information.

**Quote:** "We need trustable information, information on the parcels, to become forest or not, and manure (fertilizer, slurry, and solid manure)."

**Narrative:** Sofia Müller is a young farmer from Germany. She comes from a farming family and has been involved in the agricultural industry since she was a child. Sofia is passionate about sustainable farming practices and has a strong commitment to protecting the environment. She believes that farming can be part of the solution in reducing carbon emissions and preserving natural resources. However, Sofia often struggles with finding reliable information and resources to support her sustainable farming practices. She is frustrated by the lack of trustworthy information related to topics such as soil fertility, forest expansion, and manure management. Sofia is determined to find solutions to these challenges in order to enhance her farming practices and contribute to a more sustainable agricultural industry.

**Additional goals:**
- Accessing relevant information on farming regulations and environmental practices
- Connecting with other farmers to exchange experiences and learn about best practices

**Additional frustrations:**
- Difficulty in finding information in her local language
- Limited accessibility of digital tools due to their high cost and lack of compatibility with small farms

**Behaviour:**
- Sofia pays close attention to details and regulations related to her farm, ensuring accuracy in her work.
- She is cautious about the source of information and filters it to ensure reliability.
- Sofia is open to technological solutions and digital tools to improve her farming practices.

**Personality:**
- Sofia is detail-oriented and meticulous in attending to the details of her work.
- She has a strong sense of environmental consciousness and is committed to finding sustainable practices.
- Sofia values reliability and prefers more trustworthy and scientific sources of information.

| Themes Used in Example 1 | | |
|---|---|---|
| Goals | Information Verification and Reliability | Agricultural Education and Training |
| Frustrations | Frustration with Lack of Training and Education | Frustration with Communication and Collaboration |
| Behaviour | Trust in Information Sources | Farm Management and Monitoring |
| Personality | Attention to Detail | Sustainability |

| **Persona – Example 2** | |
|---|---|
| **Name**: Mario Rossi<br>**Country:** Italy<br>**Age:** Middle-aged | 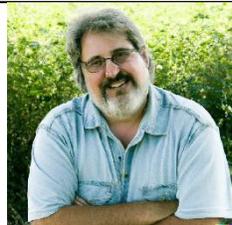 |

**Name**: Mario Rossi

**Country:** Italy

**Age:** Middle-aged

**Goal & Frustration:** Mario's main goal is to increase digitalization in his farming practices to improve efficiency and decision-making. His main frustration is the lack of user-friendly and centralized digital tools that provide all the necessary information in one place.

**Quote:** "Digital sources will become even more important in the future than they already are today."

**Narrative:** Mario is a middle-aged farmer from Italy who is passionate about sustainable agriculture and finding ways to improve his farming practices. He is open to new ideas and technologies that can help him optimize his work and reduce his environmental impact. Mario has a strong desire to learn and explore new information, regularly seeking out information online and reading magazines to stay updated on the latest agricultural practices.

**Additional goals:**
- To use digital tools for precise and efficient farming operations
- To receive market feedback for better decision-making

**Additional frustrations:**
- Limited access to information from neighboring countries due to language barriers
- Lack of visual aids, such as photos and videos, to aid in identifying diseases and other issues in farming

**Behaviour:**
- Problem-solving online: Mario actively seeks out solutions to specific problems online.
- Trust in digital tools: Mario trusts digital tools for farm management and crop monitoring, but still wants to make the final decision as a human operator.
- Information-seeking: Mario filters information to ensure it is correct and reliable, and prefers practical and illustrated content.

**Personality:**
- Curiosity: Mario has a strong desire to learn and explore new information and knowledge.
- Openness to Learning: Mario is willing to learn and adapt to new technologies and methods.
- Environmental consciousness: Mario is aware of the impact of farming on the environment and is interested in finding sustainable practices to reduce carbon emissions.

| Themes Used in Example 2 | | |
| --- | --- | --- |
| Goals | Digital Tools for Farm Management (part 1) | Agricultural Education and Training |
| Frustrations | Frustration with Limited Access to Information | Frustration with Lack of Visual Aids |
| Behaviour | Online Information Access and Behavior (part 2) | Importance of Reliable and User-Friendly Digital Tools |
| Personality | Willingness to Learn and Adapt | Sustainability |

| Persona – Example 3 | |
| --- | --- |
| **Name and surname:** João Silva<br>**Country:** Portugal<br>**Age:** Middle-aged | 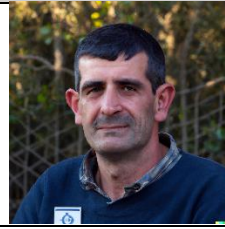 |

**Goal & Frustration:** João's main goal is to improve animal breeding processes and monitoring for his farm. His main frustration is the difficulty in finding reliable and immediate information online.

**Quote:** "If you have a problem out in the field, it is best to find a solution immediately. Otherwise, if you have to sit down again in the evening to look something up or read something, it won't work."

**Narrative:** João Silva is a middle-aged farmer from Portugal. He has been running his farm for several years and is interested in implementing digital tools to improve animal breeding processes and monitor livestock health. He is frustrated with the difficulties of finding reliable information online and often relies on offline sources, such as local meetings and journals, for information.

**Additional goals:**
- Improve efficiency and productivity through mechanization.
- Find solutions to specific problems related to livestock and farming techniques.

**Additional frustrations:**
- Difficulty in accessing information due to language barriers.
- Overload of irrelevant information when searching online.

**Behaviour:**
- Immediately seeks solutions to problems by accessing information online.
- Regularly uses specific websites for information.
- Relies on offline sources, such as local meetings and journals.

**Personality:**
- Digital literacy: Skilled in using digital tools and platforms.
- Innovative: Open to new developments in technology and digital solutions.
- Resourceful: Seeks information directly from sources and finds creative solutions.

| Themes Used in Example 3 | | |
|---|---|---|
| Goals | Animal Breeding and Livestock Management | Precision Agriculture and Crop Management |
| Frustrations | Frustration with Accessing Information Online (part 1) | Frustration with Beekeeping |
| Behaviour | Online Information Access and Behavior (part 1) | Information Gathering and Problem-Solving |
| Personality | Digital Savvy (part 1) | Attention to Detail |

| Persona – Example 4 | |
|---|---|
| **Name:** Lucia Rossi<br>**Country:** Italy<br>**Age:** Middle age | 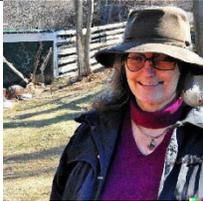 |

**Goal & Frustration:** Lucia's main goal is to stay up-to-date with the latest information and knowledge related to the agricultural sector. Her frustration lies in the lack of immediate access to information when needed.

She says, "If you have a problem out in the field, it is best to find a solution immediately."

**Narrative:** Lucia Rossi is a 45-year-old farmer from Italy. She has been running her family farm for the past 20 years and has experience in dairy farming. Lucia is passionate about organic farming and sustainability practices. She values traditional methods and believes in the importance of the farmer's expertise. Lucia focuses on staying informed about new regulations and technologies in order to optimize her farming operations and improve profitability. She relies on offline sources of information, such as local meetings with dairy farmers, agricultural journals, and TV programs on rural development. Lucia seeks advice from her family, veterinarians, and other farmers to solve problems and make informed decisions. She also values social interaction and relationships within the agricultural community, as she believes in the power of sharing experiences and exchanging information. Lucia is conscious of the impact of farming on the environment and is interested in finding ways to reduce carbon emissions.

**Additional Goals:**
- Accessing relevant information about dairy farming regulations and expansion of forest plots
- Building a network and collaborating with other farmers and industry professionals

**Additional Frustrations:**
- Difficulty typing search queries due to language barriers
- Lack of visuals in digital sources to aid in identifying diseases and other issues in farming

**Behaviors:**
- Problem-solving online by searching for solutions to specific problems
- Seeking advice from family, veterinarians, and other farmers
- Preference for visual learning and content, such as videos and diagrams

**Personality Traits:**
- Seeking advice and valuing traditional methods and relationships
- Nostalgic and appreciative of the past
- Environmental conscious and interested in finding sustainable practices

| Themes Used in Example 4 | | |
|---|---|---|
| Goals | Organic Farming and Sustainability | Farming Information and Knowledge Access (part 1) |
| Frustrations | Frustration with Accessing Information Online (part 1) | Frustration with Lack of Visual Aids |
| Behaviour | Online Information Access and Behavior (part 2) | Preferences for Visual Learning and Content |
| Personality | Traditional Values | Sustainability |

**Discussion**

This paper sought to extend work previously done by the author in performing phase 6 of Thematic Analysis conducted with LLMs, for writing-up user personas. The paper has proposed an improved prompting and process to use the results of a Thematic Analysis for creating Personas. It constitutes therefore an incremental improvement on the previous process. The impetus for this work came from some recent developments. First, the release of a LLM (GPT3.5Turbo-16k) which allows the processing of a reasonably large number of tokens, therefore allowing larger prompts, accommodating the use of several Themes, and larger responses with the production of richer Personas. A second, and more important, aspect are the reflections on the impact of LLM on UX research, and comments (presented in the introduction) on whether LLM can capture aspects such as user behaviour of personality traits of users.

We will start reflecting on the second point. Overall looking at the Themes in Table 5 and 6 in particular it does seem that the model, has some capacity to identify behaviour and personality traits of users from interview data, when instructed with a simple definition inside the prompt for the definition of initial codes. This then can lead to Themes which capture these aspects of the users. For example, captured behaviours (as Themes) include the like of how the user accesses online or offline information, collaboration with other actors (e.g. other farmers, agricultural specialists) or the use of digital tools to manage farms. Personality traits instead include the like of being a person attentive to details, being open to collaboration with others in agriculture, or being attentive to traditional values in agriculture. There are some occasional instances where perhaps the Themes do not capture entirely well either behaviour or personality traits, of which possibly the clearest example is the Theme "Problem Solving Skills" in the Personality Traits set, which the model defined as the "ability to analyze and evaluate information to make informed decisions and solve problems.". This appears possibly more a behaviour than something relating to the personality of the user, and it may be due to the definition of personality Traits used in the prompt which also hints at behaviours. In this work I did not operate Phases 4 of Thematic Analysis, which relates to revising the Themes, the "Problem Solving Skills" thus might have been dropped in phase 4, and I believe generally therefore that the fact that very few time may not match with the definition of what the model was supposed to capture, does not generally invalidate the model's capacity to perform at least satisfactorily the analysis of user interview data. As we started the paper reflecting on the observation by a leading UX scholar that LLMs may not be able to capture aspects of the user behaviour, I would suggest that LLMs can with proper prompting and good quality data identify at least some aspects of user behaviour (and personality traits) that can serve some of the work of UX researchers. It is clear that these proposed here are just some preliminary observations and more work will be needed, however.

The second aspect we need to reflect on is the use of the large GPT model supporting the processing (prompt + response) of 16 thousand tokens, and the consequent possibility to create an improved process and prompting for User Personas. The proposed workflow used 4 sets of Themes and from each of these, two Themes are then selected (randomly) to create Personas. Therefore 4 tuples (8 Themes in total) are passed inside the prompt, and the proposed prompt instructs the model about the structure of the Personas. Overall, the model does well in using the materials from the Themes to build compelling Personas. Behaviours and Personality Traits are also well presented. Although these appear as bullet points in the Personas, they are also picked up by the model in the extended narrative of the Personas. Take for example this passage from the Persona 2 (Mario) narrative where the model states that "He is open to new ideas", this is also then reflected in the separate Personality Traits list as "Openness to learning". Likewise in the Persona 4 (Lucia) the narrative describes that "She values traditional methods and believes in the importance of the farmer's expertise." and the bullet points list shows a Personality Trait related to "Nostalgic and appreciative of the past". Behaviours present a similar pattern, for instance in the Persona 3 the model picks up that João "Relies on offline sources, such as local meetings and journals.", which is also reflected in the bullet points list. Likewise in Persona 1, the model stipulates in the narrative that "Sofia often struggles with finding reliable information and resources to support her sustainable farming practices", which also is reflected in the Behaviours list. We had already observed in previous work (see De Paoli, 2023b), that the model also does well in showing a good quote for illustrating the main goal of the Personas, and this capacity is again reproduced in the Personas shown in this paper. Just as an example, we can look at Persona 2, where "Mario's main goal is to increase digitalization in his farming practices to improve efficiency and decision-making". The quote retrieved from original interview data then is as follows: "Digital sources will become even more important in the future than they already are today.".

As previously argued (De Paoli 2023b), the Personas that can be generated using LLMs and qualitative data should probably be assumed more as canvas than finished products, to be reviewed and potentially enriched by researchers/practitioners, before to be used for the UX process. Whilst workflows different than the one proposed here may be developed, the random selection of Themes from a pool of different sets of Themes, allows significant variety due to the high number of possible combinations. In this research for example 40 original Themes were produced (divided into 4 sets) and 8 are used each time, randomly selected. With this number the possible combinations $C(n,r)=C(40,8)$ well reach above 76 million, offering potential significant variety and the possibility for researchers to generate and select the initial canvas that better represent their users.

As discussed in the literature review, LLMs generated Personas, based on qualitative data and analysis could probably be seen as a middle ground between the "more traditional" qualitative Personas crafted entirely by researchers from qualitative data (and related analysis) and the data-driven personas, reusing a pool of existing data such as analytics or surveys, and often crafted with the support of algorithms. This middle ground is perhaps evident, since the LLMs generated Personas can rely on the richness of qualitative data, but at the same on the capacity of AI solutions to scale up the process of Personas creation. More work however will be required to understand how LLMs personas relate to existing literature and practice

The final point of the discussion should touch on the role of LLMs in UX research. Whilst the use of LLM for building Personas is just one potential application of these innovations that may affect UX research, it is evident there is some potential to explore. Again, the approach should be one of Human-AI collaboration where the work of researchers is enhanced and facilitated by these technologies.

## Limitations

As we observed, even using a LLM which allows the processing of 16 thousand tokens the size of certain Themes (when brought together with others in the randomised selection) would produce the max tokens limit error of the model. As said earlier, I did not write python code to overcome the error (which could have simply entailed a pre-check of the length of the combined tuples), as my goal here was demonstrative in terms of producing a richer set of Personas from an LLM drive TA. However, should a similar type of workflow be adopted in production, then avoiding from the outset the error will become important. One recommendation would be to establish a max length for each of the generated Themes, so that each random composition of tuple can be processed by the model without this error.

## Conclusion

This draft manuscript has presented and discussed an improved prompting and process for creating User Personas with an LLM, based on the results of a TA of qualitative interviews also performed with the same LLM. A workflow for creating user personas has been discussed and the results of the analysis and a set of Personas were articulated, presented and discussed. Whilst this is just an incremental contribution to previous work, it offers a further step toward understanding LLMs potential use in research in the social sciences.

# Appendix – Additional examples of Personas

| |
|---|
| **Further example 1** |
| **Name:** Ana Popescu<br>**Country:** Romania<br>**Age:** Middle-aged<br><br>**Goal & Frustration:** Ana's main goal is to access reliable information and resources in beekeeping to improve her practices. Her main frustration is the lack of reliable sources and the need for necessary digital skills.<br><br>**Quote:** "They don't have good digital skills. And so everything should be very intuitive."<br><br>**Narrative:** Ana Popescu is a middle-aged beekeeper from Romania. She has been passionate about beekeeping since she was young and has dedicated her life to this profession. She manages her own bee farm and is continuously looking for ways to improve her practices and expand her knowledge. Ana believes that access to reliable information is crucial for the success of her beekeeping business.<br><br>**Additional goals:**<br>1. Networking and collaboration with other beekeepers to exchange experiences and knowledge.<br>2. Staying up-to-date with the latest developments and technologies in beekeeping.<br><br>**Additional frustrations:**<br>1. Difficulty in finding accurate and relevant information for her specific needs.<br>2. Lack of digital literacy and necessary digital skills to effectively use digital tools.<br><br>**Behaviour:**<br>1. Ana actively searches for reliable sources and verifies information before applying it in her practices.<br>2. She is meticulous and pays close attention to details in her beekeeping practices.<br>3. Ana prefers to collaborate with other beekeepers and industry experts to exchange experiences and gain knowledge.<br><br>**Personality:**<br>1. Detail-oriented: Ana pays close attention to small details and ensures accuracy in her practices.<br>2. Problem-solving: She actively seeks solutions to challenges in her beekeeping business.<br>3. Strategic thinking: Ana analyzes and evaluates information to make informed decisions for her bee farm. |

| Themes Used in Further example 1 | | |
|---|---|---|
| Goals | Language and Localization | Organic Farming and Sustainability |
| Frustrations | Frustration with Beekeeping | Frustration with Accessing Information Online |
| Behaviour | Trust in Information Sources | Farm Management and Monitoring |
| Personality | Attention to Detail | Problem-solving Skills |

| Further example 2 |
|---|
| **Name:** Maria Lopez<br>**Country:** Spain<br>**Age:** Middle age<br><br>**Goal & Frustration:**<br>**Goal:** Maria's main goal is to access trustworthy information related to parcels, forest expansion, and manure in order to make informed decisions and improve her farming practices.<br>**Quote:** "We need trustable information, information on the parcels, to become forest or not, and manure (fertilizer, slurry, and solid manure)."<br><br>**Narrative:** Maria Lopez, a middle-aged farmer from Spain, is deeply passionate about agriculture and the environment. She runs a small farm where she practices sustainable farming methods and strives to find ways to reduce carbon emissions. Maria is frustrated by the lack of clear and reliable information regarding parcel management, forest expansion, and manure. She wants to make informed decisions about these practices, but finding trustworthy information has been a challenge. Maria is dedicated to staying informed about the latest developments in the agricultural sector and is always seeking innovative solutions to improve her farming operation. She is open to using digital tools and technologies to make her work easier and more efficient, but she values the final human decision-making process. Maria believes that digital sources and platforms play a crucial role in providing information to farmers and fostering collaboration among them. She is also concerned about the resistance to change in traditional agriculture practices and hopes to see more farmers embrace digital tools and new methods. Maria is eager to collaborate with other farmers and the Farmbook team to share knowledge and best practices.<br><br>**Additional goals:**<br>- Access reliable information on animal breeding and problem-solving in livestock farming.<br>- Stay up-to-date with the latest developments in organic farming, regulations, and practices.<br><br>**Additional frustrations:**<br>- Limited international market access due to protectionism.<br>- Seasonal gaps in production that require additional efforts to fill. |

**Behaviour:**
- Uses digital tools and platforms regularly to search for information and stay informed.
- Prefers visual and illustrated content for easy comprehension.
- Filters out irrelevant or unreliable information to find trustworthy sources.

**Personality**:
- Environmentally conscious with a focus on finding sustainable practices.
- Digitally literate and open to new technologies and innovations.
- Resourceful and innovative in seeking solutions to farming challenges.

| Themes Used in Further example 2 | | |
|---|---|---|
| Goals | Farming Information and Knowledge Access | Organic Farming and Sustainability |
| Frustrations | Frustration with Agricultural Practices and Regulations | 'Frustration with Language Barriers |
| Behaviour | Online Information Access and Behavior (part 1) | Trust in Information Sources |
| Personality | Sustainability | Digital Savvy (part 1) |

| **Further example 3** |
|---|
| **Name:** Maria Silva <br> **Country:** Portugal <br> **Age:** Middle age <br><br> **Goal & Frustration:** Maria's main goal is to access trustworthy and reliable information related to farming practices and regulations. Her frustration is the difficulty of finding relevant and accurate information online, as there is an abundance of irrelevant or misleading information. <br><br> **Quote:** "The strongest are the government's portals which have good information and real." <br><br> **Narrative:** Maria is a middle-aged farmer from Portugal. She owns a small farm where she raises livestock and cultivates crops. Maria is passionate about sustainable farming practices and is always looking for ways to improve efficiency and reduce the environmental impact of her farm. She values practical experience and learning from other farmers. However, she often relies on online sources to gather information and stay informed about the latest farming practices and regulations. Maria finds it frustrating to encounter poorly maintained websites, language barriers, and an overwhelming amount of information that may not be relevant or reliable. She believes that accessing trustworthy information is crucial for making informed decisions and improving her farming practices. <br><br> **Additional goals:** <br> - To collaborate and exchange information with other farmers and industry professionals. <br> - To stay up-to-date with the latest innovations and technologies in farming. |

**Additional frustrations:**
- Difficulty filtering out misleading information found online.
- Lack of user-friendly platforms and tools for accessing agricultural information.

**Behaviour:**
- Uses a smartphone as an essential tool for work and research.
- Relies on both online and offline sources to gather information.
- Prefers interactive and visual learning tools such as videos and images.

**Personality:**
- Attention to detail
- Environmental consciousness
- Concern for the use of sustainable farming practices

| Themes Used in Further example 3 | | |
|---|---|---|
| Goals | Language and Localization | Farming Information and Knowledge Access |
| Frustrations | Frustration with Accessing Information Online (part 1) | Frustration with Agricultural Practices and Regulations |
| Behaviour | Use of Digital Tools for Specific Purposes | Information Gathering and Problem-Solving |
| Personality | Attention to Detail | Sustainability |

| Further example 4 |
|---|
| **Name and surname:** Maria Rossi<br>**Country:** Italy<br>**Age:** Middle age<br><br>**Goal & Frustration:** Maria's main goal is to incorporate digital tools into her farm management practices for precision agriculture and crop management. She is frustrated with the lack of reliable and easy-to-use digital tools that provide accurate information and support for her farm operations.<br><br>**Quote:** "Digital sources will become even more important in the future than they already are today."<br><br>**Narrative:** Maria is a middle-aged farmer in Italy who owns a small farm that specializes in both beef cattle and arable farming. She has a strong passion for sustainable farming practices and is committed to finding innovative solutions to improve her farm's productivity while minimizing the environmental impact. Maria is open to incorporating digital tools into her |

farm management practices, but she has been struggling to find reliable and user-friendly options that suit her specific needs. She often searches for information online in English, as she believes it provides more in-depth answers and access to a wider range of resources. Maria also values networking and collaboration with other farmers and experts in the field, attending conferences and exchanging knowledge. She is actively looking for tools and platforms that encourage collaboration and information sharing among farmers. Despite her interest in technological solutions, Maria also relies on her own experience and the input of experts to make informed decisions for her farm. She prioritizes trustable information sources, such as governmental institutions and scientific organizations, and is skeptical of information found on social media platforms and forums. Maria is committed to the long-term sustainability of her farm and is motivated to find effective and practical digital tools that can support her in achieving her goals.

**Additional goals:**
- Accessing reliable and local language resources for farm management
- Early identification of crop issues

**Additional frustrations:**
- Difficulty in finding practical and trustworthy digital tools for agro-technical operations
- Lack of user-generated content and collaboration platforms specific to agriculture

**Behaviour:**
- Maria actively seeks information online in different languages based on the depth of the answers and the source of information.
- She prefers reliable sources of information and filters out misleading or irrelevant information.
- Maria values collaboration and networking with other farmers and experts in the field.

**Personality:**
- Maria is environmentally conscious and prioritizes sustainable practices.
- She is willing to work with others and share knowledge and expertise to achieve common goals.
- Maria is open-minded and adaptable to new technologies and digital tools.

| Themes Used in Further Example 4 | | |
| --- | --- | --- |
| Goals | Agricultural Education and Training | Precision Agriculture and Crop Management |
| Frustrations | Frustration with Lack of Visual Aid | Frustration with Communication and Collaboration |
| Behaviour | Trust in Information Sources | Language Preferences and Information Source |
| Personality | Sustainability | Collaboration and Information Sharing |

**Further example 5**

**Name:** Anna Schmidt
**Country:** Germany
**Age:** Middle age

**Goal & Frustration:** Anna's main goal is to access information from neighboring countries, such as Bulgaria and Hungary, to improve her farming practices. Her main frustration is the lack of access to information due to language barriers and limited information sharing.

**Quote:** "But I don't know what is happening in neighboring countries. Like Bulgaria or Hungary, which have maybe similar conditions to mine. Even Serbia is not easy access as far as information and maybe something that is happening in Germany or in Denmark could be of interest for me also."

**Narrative:** Anna Schmidt is a 45-year-old farmer from Germany. She comes from a family with a long history of farming and has been working on her own farm for over 20 years. Anna is passionate about agriculture and is always looking for ways to improve her farming practices. However, she faces the challenge of limited access to information, especially from neighboring countries with similar conditions.

Anna spends her free time attending agricultural fairs and reading magazines to stay informed about the latest trends and technologies in farming. She has a good relationship with her local veterinarian, who provides valuable insights and knowledge. However, Anna believes that there is a wealth of information available from other farmers and scientific institutions in neighboring countries.

Anna is an innovative and resourceful farmer. She has developed her own web applications and uses digital tools to monitor her crops and make data-driven decisions. Anna is open to using new technologies and is willing to learn and adapt to them. She values efficiency and believes that digital tools can greatly improve farm management practices.

**Additional goals:**
- Stay informed about the latest advancements in animal breeding techniques
- Find solutions to manage fauna in the area

**Additional frustrations**:
- Limited availability of information due to the small market and lack of big companies in Germany
- Difficulty in understanding complicated digital tools due to lack of time and knowledge

**Behaviour:**
- Seeks information directly from reliable sources, such as scientific institutions and manufacturers
- Uses digital tools and online resources effectively to gather information
- Relies on her own experience and experimentation to find solutions to problems

**Personality:**

- Curious and always eager to learn and explore new information and knowledge
- Open-minded and willing to embrace new ideas and technologies
- Resourceful and able to find creative solutions to problems.

| Themes Used in Further Example 5 | | |
|---|---|---|
| Goals | Specific Farming Practices and Topics | Precision Agriculture and Crop Management |
| Frustrations | Frustration with Limited Access to Information | Frustration with Lack of Training and Education |
| Behaviour | Trust in Information Sources | Farm Management and Monitoring |
| Personality | Willingness to Learn and Adapt | Digital Savvy |